\documentclass[aip,apl,reprint,graphicx,%
 amsmath,amssymb,
]{revtex4-1}
\usepackage{graphicx}
\usepackage{dcolumn}
\usepackage{bm}
\draft 

\begin{document}


\title{Inverse Spin Hall Effect in Ferromagnetic Metal with Rashba Spin Orbit Coupling} 

\author{M.-J. Xing}
\affiliation{Computational Nanoelectronics and Nano-device Laboratory, Electronic and Computer Engineering Department, National University of Singapore, Singapore, 117576}
\affiliation{State Key Laboratory for Advanced Metals and Materials, School of Materials Science and Engineering, University of Science and Technology Beijing, China, 100083}
\author{M. B. A. Jalil}
\email{elembaj@nus.edu.sg}
\affiliation{Computational Nanoelectronics and Nano-device Laboratory, Electronic and Computer Engineering Department, National University of Singapore, Singapore, 117576}
\affiliation{Information Storage Materials Laboratory, Electronic and Computer Engineering Department, National University of Singapore, Singapore, 117576}
\author{Seng Ghee Tan}
\affiliation{Computational Nanoelectronics and Nano-device Laboratory, Electronic and Computer Engineering Department, National University of Singapore, Singapore, 117576}
\affiliation{Data Storage Institute, Agency for Science, Technology and Research (A*STAR). DSI Building, 5 Engineering Drive 1, Singapore, 117608}
\author{Y. Jiang}
\affiliation{State Key Laboratory for Advanced Metals and Materials, School of Materials Science and Engineering, University of Science and Technology Beijing, China, 100083}

\date{\today}

\begin{abstract}
We report an intrinsic form of the inverse spin Hall effect (ISHE) in ferromagnetic (FM) metal with Rashba spin orbit coupling (RSOC), which is driven by a normal charge current. Unlike the conventional form, the ISHE can be induced without the need for spin current injection from an external source. Our theoretical results show that Hall voltage is generated when the FM moment is perpendicular to the ferromagnetic layer. The polarity of the Hall voltage is reversed upon switching the FM moment to the opposite direction, thus promising a useful readback mechanism for memory or logic applications.
\end{abstract}


\maketitle 

\section{Introduction}
Recent research showed that Rashba spin orbit coupling (RSOC) at the surfaces of metals can be enhanced by
the presence of heavy atoms \cite{Gambardella,Ast} and/or surface oxidation \cite{LaShell} in adjacent layers. This interfacial enhancement
enables significant RSOC effect to be manifested in ferromagnetic metals with small or moderate atomic number at room
temperature \cite{Miron}, whereas previously, strong RSOC effect is confined only to semiconductor heterostructures.
In this paper, we choose a typical ferromagnetic (FM) metal (Co) as the central conducting layer, sandwiched between
an oxide and a Pt layer, the latter supplying the heavy atoms [see Fig. 1(a)]. The electron accumulation which develops
in the FM layer in the presence of a charge (unpolarized) current is theoretically evaluated via the non-equilibrium
Green's function (NEGF) method in the ballistic limit. In the presence of $s$-$d$ coupling, the incoming charge current
becomes polarized by the FM moments in the central region. When this intrinsic polarization of current is coupled to
the RSOC, an inverse spin Hall effect (ISHE) \cite{saitoh,Hankiewicz} will be induced. Thus, a Hall voltage is generated without the need
for spin injection from an external spin polarizing layer. By contrast, in previous works, the ISHE is experimentally
realized by injecting spin polarized current \cite{Valenzuela,Zhang} from an external FM electrode, or by the inflow of pure spin
current \cite{Hankiewicz,Saitoh1,Kimura,Xing,Li,Ando}, generated externally e.g. via spin pumping or non-local spin accumulation. In this work, we show
theoretically that Hall voltage can be generated when the FM moment in the central region is oriented perpendicular to
the plane, which persists at room temperature. Furthermore, the generated Hall voltage can be reversed symmetrically
when the FM moment is switched to the opposite direction. Thus, the charge current-induced ISHE signal can be used to
detect the polarity of the FM moment, and potentially serve as a read-back mechanism in memory applications.
\begin{figure}[t]
  \centering
   \includegraphics[width=6.0cm]{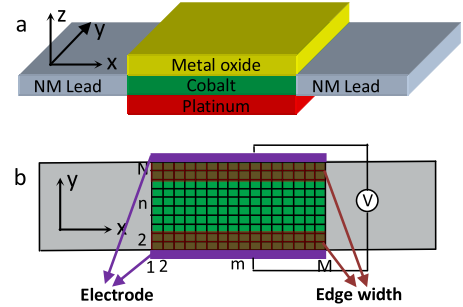}
     \caption{(a) Schematic diagram of the proposed FM moment detector utilizing the ISHE phenomenon. (b) Lattice discretization of the device for tight-binding NEGF calculation (top view). The Hall voltage $V_t$ is the potential difference between the two strip electrodes running along the top and bottom edges of the central region. The covered area schematically shows the edge width which will be used for the Hall voltage calculation.}
    \label{fige1}
\end{figure}

\section{Model Hamiltonian and Theory}
The schematic diagram of the FM moment detector is shown in Fig. 1(a); the central region comprises of a triple-layer structure for the enhancement of RSOC within the FM (Co) layer \cite{Miron}. Charge accumulation within the Co layer is calculated via tight-binding NEGF method \cite{Xing}. To perform the tight-binding calculation, the central region of the device is discretized into $(M\times N)$ lattice of points [see Fig. 1(b)]. The conduction electrons within the Co layer experiences the RSOC effect and the $s$-$d$ exchange interaction with the local FM moments $\textbf{M}(\theta,\phi)$. Thus, the Hamiltonian of the central region can be expressed as $H_{C}=H_{K}+H_{M}+H_{Rso}$, where $H_{K}$ is the kinetic term, $H_{M}$ the $s$-$d$ coupling term, and $H_{Rso}$ the RSOC term. The Hamiltonian can be expressed as \cite{Rashba}:
\begin{eqnarray}
H_{K}&=&\sum_{mn\sigma}[4td^{\dagger}_{mn\sigma}d_{mn\sigma}\nonumber\\
&-&t(d^{\dagger}_{m+1,n\sigma}d_{mn\sigma}+d^{\dagger}_{m,n+1\sigma}d_{mn\sigma}+\mathrm{h.c.})],\\\label{eq5}
H_{M}&=&\sum_{mn\sigma}\mathrm{sgn}[\sigma] M\cos(\theta)d^{\dagger}_{mn\sigma}d_{mn\sigma}\nonumber\\
& &+M \sin(\theta)e^{i\sigma\phi}d^{\dagger}_{mn\sigma}d_{mn\bar{\sigma}},\\\label{eq2}
H_{Rso}&=&\sum_{mn\sigma\sigma'}-it_{so}[(d^{\dagger}_{m+1,n}d_{mn}-d^{\dagger}_{m-1,n}d_{mn})\otimes\hat{\sigma}_{y}\nonumber\\
& &-(d^{\dagger}_{m,n+1}d_{mn}-d^{\dagger}_{m,n-1}d_{mn})\otimes\hat{\sigma}_{x}],\label{eq4}
\end{eqnarray}
where $M$ denotes the $s$-$d$ coupling strength, $t_{so}=\frac{\alpha}{2a}$ denotes the RSOC strength. Similarly, the Hamiltonian of the normal metal (NM) leads, and the coupling energy between the leads and central region can be expressed as:
\begin{eqnarray}
H_{L(R)}&=&\sum_{mn\sigma}[4ta^{\dagger}_{mn\sigma}a_{mn\sigma}\nonumber\\
&-&t(a^{\dagger}_{m+1,n\sigma}a_{mn\sigma}+a^{\dagger}_{m,n+1\sigma}a_{mn\sigma}+\mathrm{h.c.})].\\\label{eq6}
H_{T}&=&\sum_{n\sigma}[t'_{L}a^{\dagger}_{0n\sigma}d_{1n\sigma}+t'_{R}a^{\dagger}_{M+1,n\sigma}d_{Mn\sigma}+\mathrm{h.c.}].\label{eq7}
\end{eqnarray}
From the eigenvalue equation of
the total Hamiltonian and the definition of retarded Green's function, one can obtain an equation: $(E-H_{mn}+i\eta)G^{n,n}_{m,m}(\sigma\sigma)=I$. From this relation, one can obtain a series of linear equations involving $G^{n,n}_{m,m}(\sigma\sigma)$ by considering each spatial point $(m,n)$. For instance, within the central region, i.e. $1<m<M$, one obtains:
\begin{widetext}
\begin{eqnarray}
I&=&[E-4t-\sigma M\cos(\theta)]G^{n,n}_{m,m}(\sigma\sigma)-e^{i\sigma\phi}M \sin(\theta)G^{n,n}_{m,m}(\bar{\sigma}\sigma)
+t[G^{n,n}_{m-1,m}(\sigma\sigma)+G^{n,n}_{m+1,m}(\sigma\sigma)\nonumber\\
&+&G^{n-1,n}_{m,m}(\sigma\sigma)+G^{n+1,n}_{m,m}(\sigma\sigma)]+t_{so}[\sigma
G^{n,n}_{m+1,m}(\bar{\sigma}\sigma)-iG^{n+1,n}_{m,m}(\bar{\sigma}\sigma)-\sigma
G^{n,n}_{m-1,m}(\bar{\sigma}\sigma)+iG^{n-1,n}_{m,m}(\bar{\sigma}\sigma)].
\label{eq8}
\end{eqnarray}
\end{widetext}
Collectively, all these equations can be
expressed in matrix form: $(E[I]-[H])[G]^{r}=I$. The infinitely large matrix $[H]$ consists of sub-matrices denoting the
Hamiltonian of the central region ($[H_{C}]$) and the coupling
coefficients between the two leads and the central region separately
($[\tau_{L/R,C}]$). Following standard procedures in the tight-binding method, one then obtains:
\begin{equation}
(E[I]-[H_{C}]-[\Sigma]^{r}_{L}-[\Sigma]^{r}_{R})[G_{C}]^{r}=I,\label{eq9}
\end{equation}
in which the non-zero terms of the self energy are:
$[\Sigma^{n,n'}_{m,m}]^{r}_{L(R)}=[\tau]^{n,n}_{m,0(M+1)}[g^{r}]^{n,n'}_{0(M+1),0(M+1)}[\tau]^{n',n'}_{0(M+1),m}$. The retarded Green's functions of the isolated left(right) lead $[g^{r}]^{n,n'}_{0(M+1),0(M+1)}$ can be expressed as (for the left lead):
\begin{equation}
[g^{r}]^{n,n'}_{0,0}=-\frac{1}{t}\sum_{i}\chi_{i}(p_{n})e^{ik_{i}a}\chi_{i}(p_{n'}).\label{eq10}
\end{equation}
In the above, $k_i$ is the wave vector along the semi-infinite longitudinal direction, $\chi_{i}(p_{n})$ is the $\tilde{i}$th eigenfunction in the transverse dimension at site $(0,n)$ in the lead, which can be expressed as:
\begin{equation}
\chi_{i}(p_{n})=\sqrt{\frac{2}{N+1}\sin{\frac{i\pi n}{N+1}}}.\label{eq11}
\end{equation}
The retarded Green's function of the central region can then be solved by inverting Eq. \eqref{eq9}. Thus one can express the lesser Green's function $[G]^{<}$ via the Langreth formula: $[G]^{<}=[G]^{r}[\Sigma]^{<}[G]^{a}$, in which
$[\Sigma]^{<}=\sum_{\mu=L,R}([\Sigma]^{a}_{\mu}-[\Sigma]^{r}_{\mu})f_{\mu}$, with $f_\mu$ being the Fermi distribution function within lead
$\mu$. The total charge accumulation for a given cell at lattice coordinate $(m,n)$ with an area of $a^2$ is given by:
\begin{equation}
\tilde{\rho}_{m,n}=-\frac{ie}{2\pi}\int^{\infty}_{-\infty}Tr[G]^{<}_{mn,mn}(E)dE,\label{eq12}
\end{equation}
where the trace is over the spin degree of freedom. The surface charge density $\rho_{m,n}$ is then given by $\rho_{m,n}=\frac{\tilde{\rho}_{m,n}}{a^2}$.

The Hall voltage at longitudinal position $x=m$ ($V_{tm}$) is given, up to a proportionality constant, by
the difference in the surface charge density between the top and bottom edges corresponding to $x=m$, i.e. $V_{tm}\propto\Delta\rho_m=\rho_{tm}-\rho_{bm}$.
In calculating the charge densities $\rho_{tm}$ and $\rho_{bm}$, we consider a finite width of each edge. The values of $\rho_{tm}$ and $\rho_{bm}$ are averaged
over some number of rows ($W$) adjacent to the top and bottom edges, such that $Wa\approx$ 0.3 nm. Thus, the surface charge density difference along $x$-direction is given by:

\begin{equation}
\Delta\rho_m=\frac{\sum^{W}_{n=1}\rho_{m,N-n+1}-\rho_{m,n}}{W}.\label{eq13}
\end{equation}
In practice, the Hall voltage $V_t$ is given by the potential difference between the two electrodes. We assume that each electrode runs along the entire length of the edges (i.e., from $m=1$ to $M$). Thus, computationally, the Hall voltage $V_t$ is given by the surface charge density difference averaged over the longitudinal dimension, i.e.,
\begin{equation}
\Delta\rho_{av}=\frac{\sum^M_{m=1}\sum^{W}_{n=1}\rho_{m,N-n+1}-\rho_{m,n}}{M\times W}.\label{eq14}
\end{equation}
\section{Results and Discussion}
The following parameters are assumed in the numerical calculations: (i) The device is modeled at room temperature ($T=300$ K); the Fermi energy of the central FM
layer is set to $7.38$ eV, which is a typical value for Co \cite{Wawrzyniaka}. (ii) The lattice cell dimension is set to $a=0.045$ nm, which is
significantly smaller than the Fermi wavelength $(a\sim\lambda/10)$,
so that the lattice Green's function model can simulate a continuum
system to a good approximation. (iii) The coupling strength is
$t=\frac{\hbar^{2}}{2ma^{2}}=18.69$ eV, while for simplicity, the
coupling between the lead and the central region is set to
$t_{L/R}=0.8t$. (iv) The RSOC strength in Eq. \eqref{eq4} is given by
$t_{so}=\frac{\alpha}{2a}$. For a typical FM RSOC material, $\alpha$ lies between $4\times10^{-11}$ and $3\times10^{-10}$ eVm
\cite{Ast,Henk,Krupin}, which translates to a range of coupling parameter values of $0.4<t_{so}<3.3$ eV.
(v) The $s$-$d$ exchange energy is set to $|M|=0.85$ eV \cite{Wakoh}. (vi) The
electrochemical potentials of the two leads are set to
$\mu_{L}=-\mu_{R}=2$ eV. (vii) Finally, the central region is
discretized into a square lattice of $(M\times N)=(200\times 100)$ of unit cells. This corresponds to an actual dimensions of (9 nm$\times$ 4.5 nm) for the central region.

\begin{figure}[t]
\centering
\includegraphics[width=4.0cm]{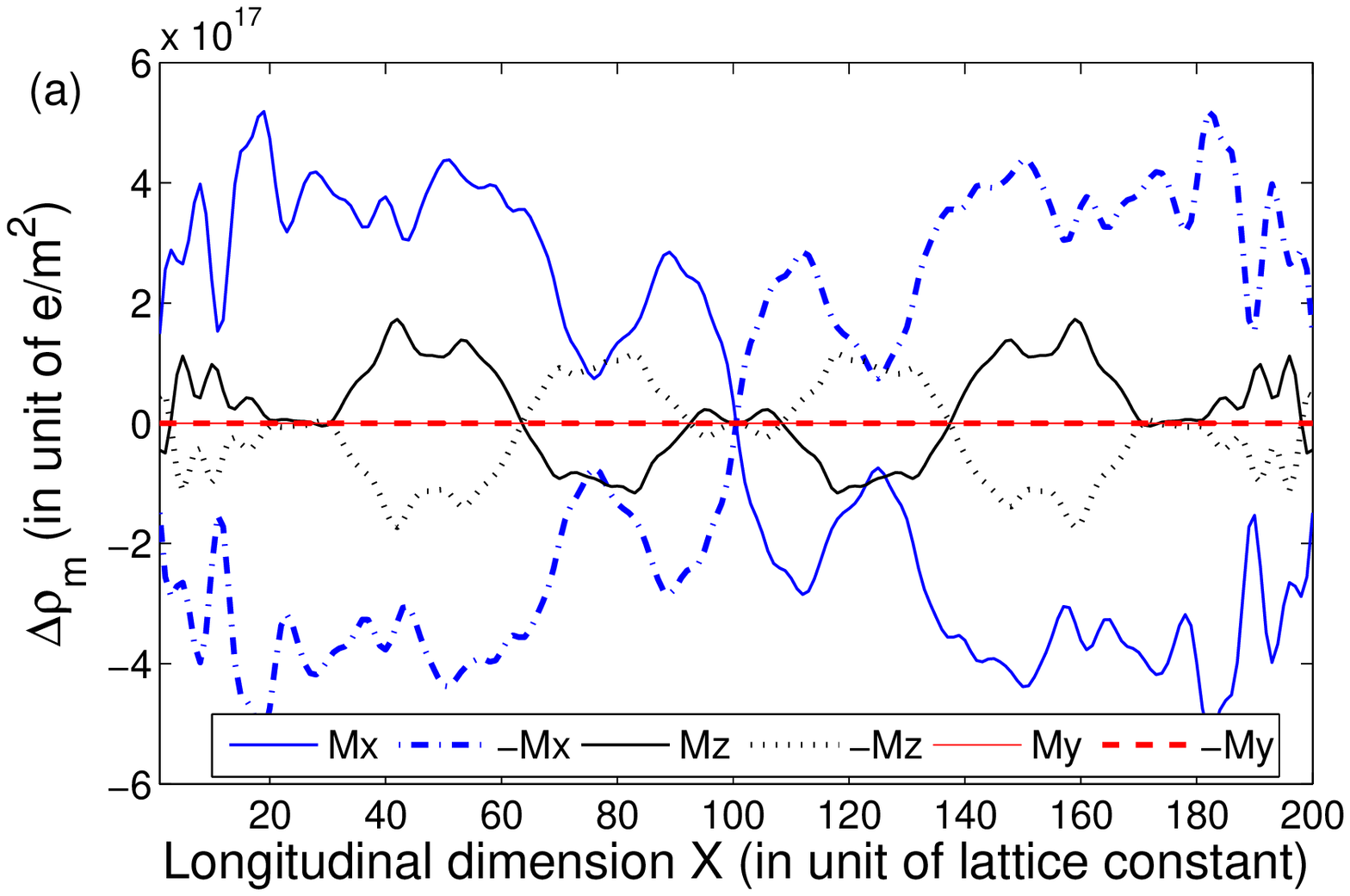}
\includegraphics[width=4.0cm]{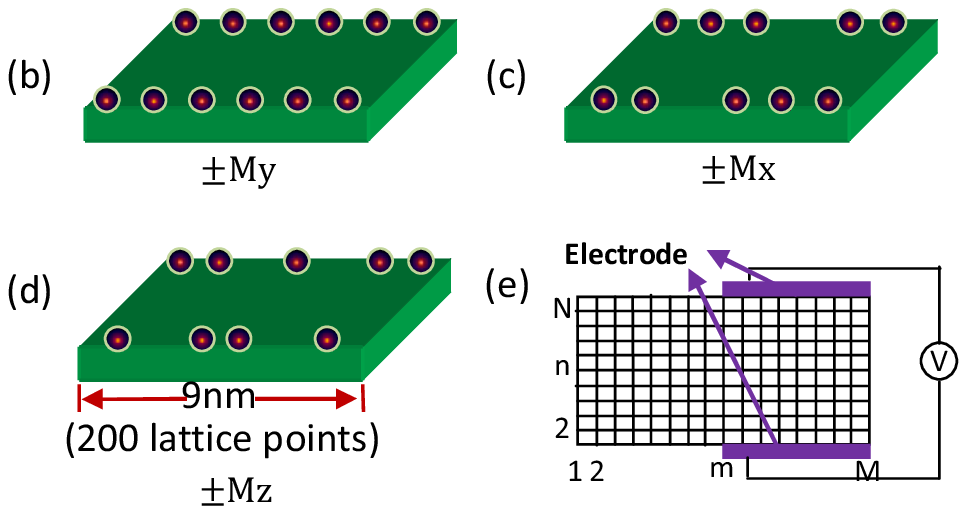}
\caption{(a) Distribution of $\Delta\rho_m$ along the longitudinal
$x$-axis. The FM moments in the central region are oriented along
the $\pm x$ (blue), $\pm y$ (red) and $\pm z$ (black) directions.
The corresponding electron density distributions are schematically
depicted in (b), (c) and (d), respectively. Fig (e) shows a possible
electrode configuration to detect the Hall voltage when FM moment is
along $\pm x$ directions. The following parameter values are
assumed: RSOC strength of $t_{so}=3$ eV, bias voltage of $V=4$ eV,
and $s$-$d$ coupling strength of $M=0.85$ eV. The central region is discretized into a
lattice of $(200\times100)$ unit cells.} \label{fige2}
\end{figure}

We first calculate the transverse charge density difference of
$\Delta\rho_m$ as a function of the longitudinal position $x$, when
the FM moments in the Co layer are separately oriented along $\pm
x$, $\pm y$ and $\pm z$ directions. The results are
plotted in Figure \ref{fige2}(a). Our results show that when the FM
moments are in the $y$-direction, the spatial charge distribution
$\rho_{m,n}$ is symmetric about the central longitudinal axis of
$n=(N+1)/2$, resulting in $\Delta\rho_m=0$. When the FM moments are
switched to $-y$ direction, the charge distribution $\rho_{m,n}$
remains symmetric about the central longitudinal axis, and hence
$\Delta\rho_m$ is still $0$ [see Fig. \ref{fige2}(b) for a schematic
representation]. Thus, when the FM moments of the Co layer are
aligned along $\pm y$, no Hall voltage would be observed. By
contrast, when the FM moments are oriented along the $x$-direction,
$\Delta\rho_m$ is symmetric about the central point
$(m,n)=((M+1)/2,(N+1)/2)$ [see Figs. \ref{fige2}(a) and (c)].
Furthermore, the sign of $\Delta\rho_m$ is reversed when the FM
moments are switched to the $-x$ direction. However, when averaged
along the entire edge, the surface charge density difference will be
$\Delta\rho_{av}=0$ due to the point symmetry. However, if the Hall
electrodes extend to only half the entire length of the top and
bottom edges [as shown schematically in Fig. \ref{fige2}(e)], a
finite Hall voltage can still be detected.

Of greater interest is the case where the FM moments are along the out-of-plane
$z$-direction. The charge density difference $\Delta\rho_m$ is symmetric about the
central vertical axis, i.e. $m=(M+1)/2$. Thus, a finite $\Delta\rho_{av}$, i.e., a Hall
voltage $V_t$ is generated [see Figs. \ref{fige2}(a) and (d)]. Since only charge current but not
spin current is injected into the system, the above can be regarded as a charge
current-induced ISHE in FM metal with RSOC.

\begin{figure}[t]
\centering
\includegraphics[width=6.0cm]{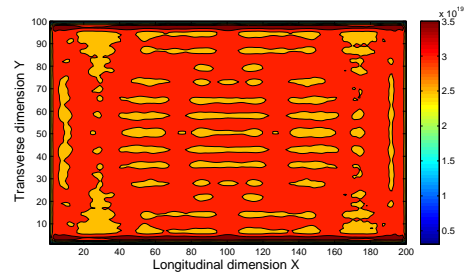}
\caption{The spatial distribution of $\rho_{m,n}$ (in unit of
$\frac{e}{m^{2}}$). The longitudinal and transverse dimensions are expressed in unit of the lattice constant $a$. The $s$-$d$ coupling strength is $M=0.85$ eV, the RSOC strength is $t_{so}=3$
eV, the bias voltage is $V=4$ eV, the central region is a $(200\times100)$ lattice ($9$ nm $\times4.5$ nm).} \label{fige3}
\end{figure}

\begin{figure}[t]
\centering
\includegraphics[width=4.0cm]{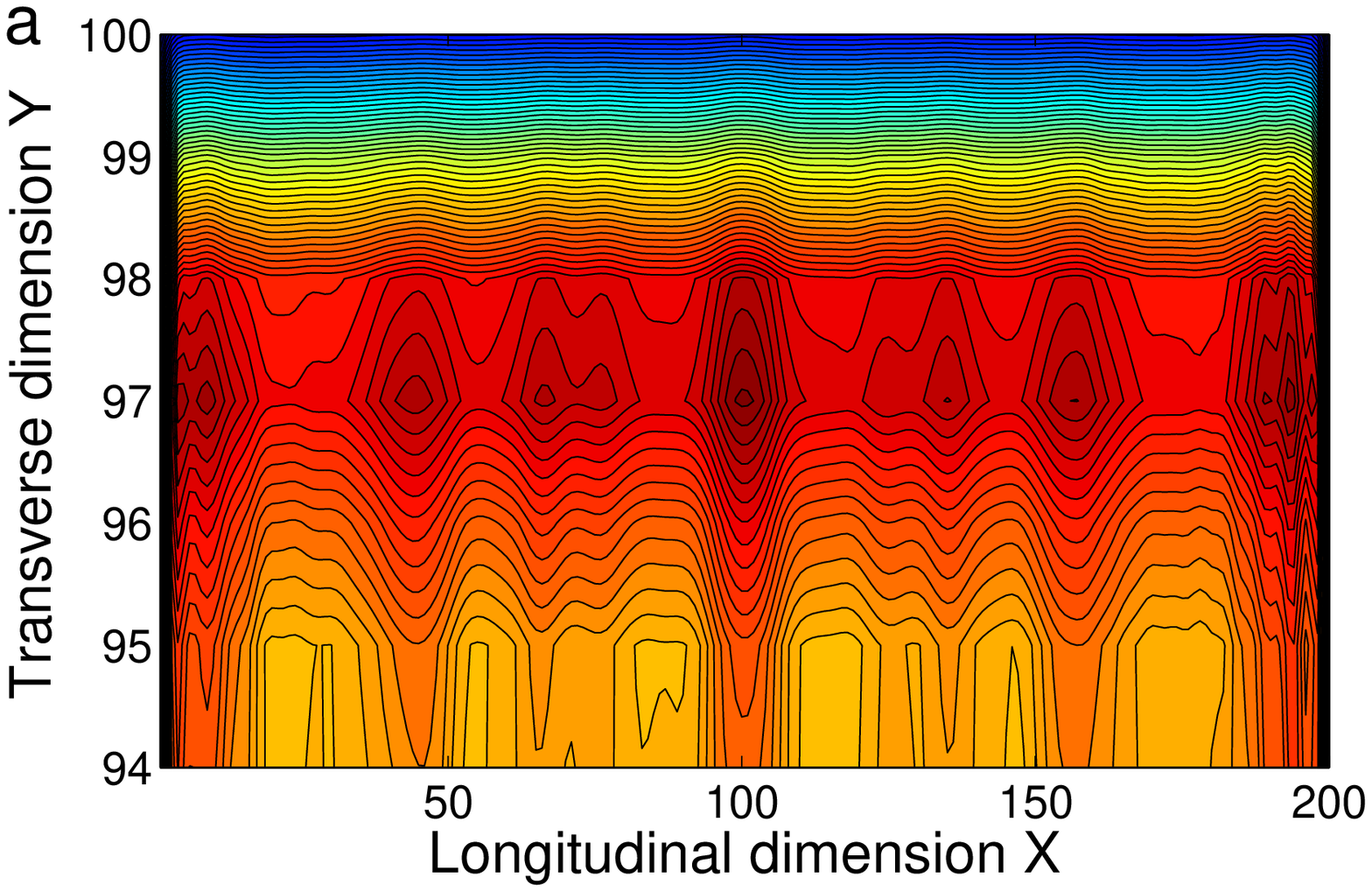}
\includegraphics[width=4.0cm]{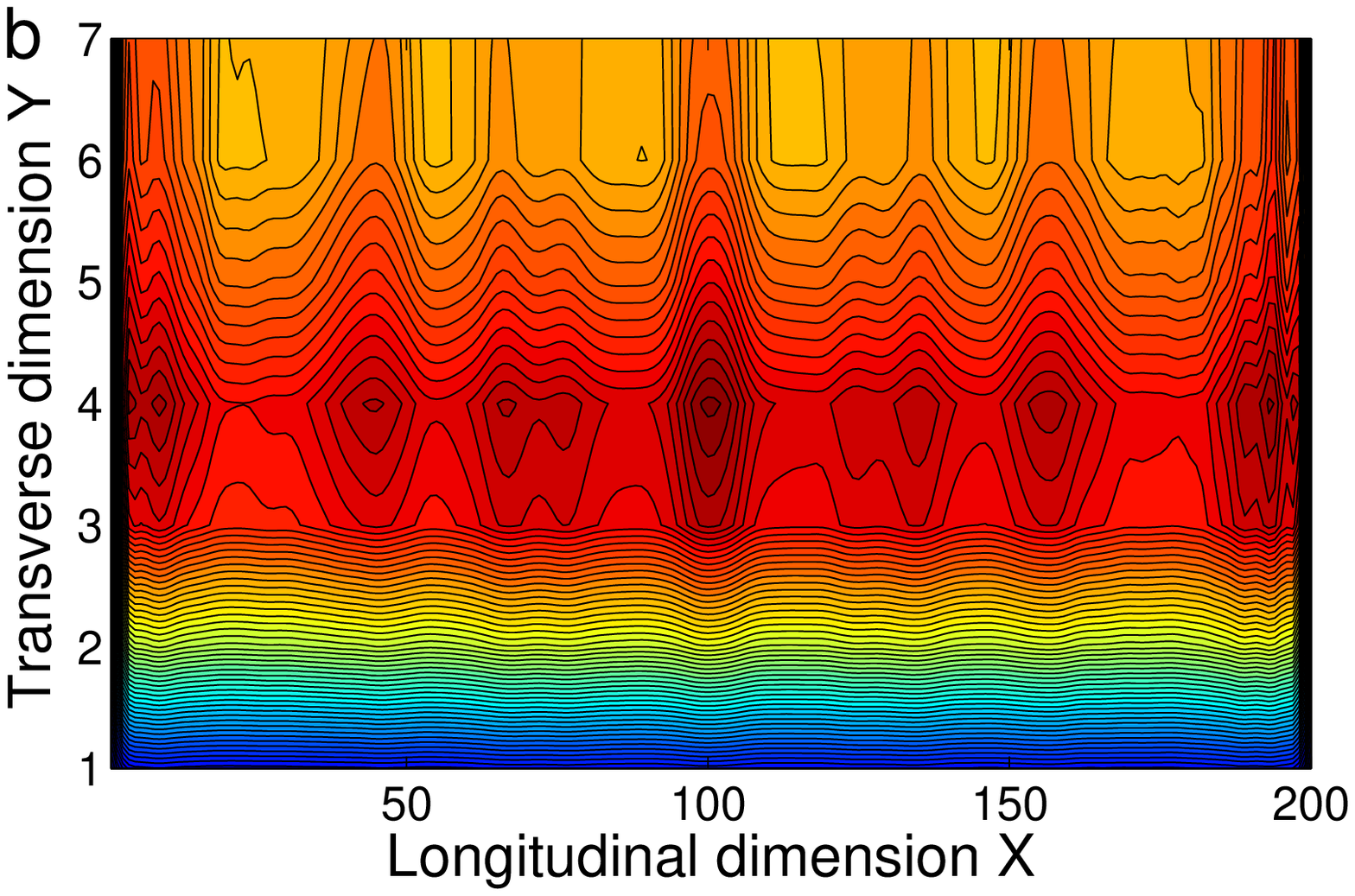}
\caption{The detailed distribution of $\rho_{tm}$ and $\rho_{bm}$ (in unit of
$\frac{e}{m^{2}}$) at (a) the top, and (b) bottom edges of Fig. 3. The longitudinal and transverse dimensions are expressed in unit of the lattice constant $a$.} \label{fige4}
\end{figure}

In the following, we will investigate this charge current-driven ISHE in greater detail.  The spatial distribution of $\rho_{m,n}$ is plotted over the central region when the FM moments are in the $+z$-direction [see Fig. 3]. For clarity, the detailed distribution of charge densities $\rho_{tm}$ and $\rho_{bm}$ are shown in Figs. 4(a) and (b). It is observed that the surface charge density is larger along the top edge, which will result in a finite Hall voltage or ISHE effect.

\begin{figure}[t]
\centering
\includegraphics[width=4.0cm]{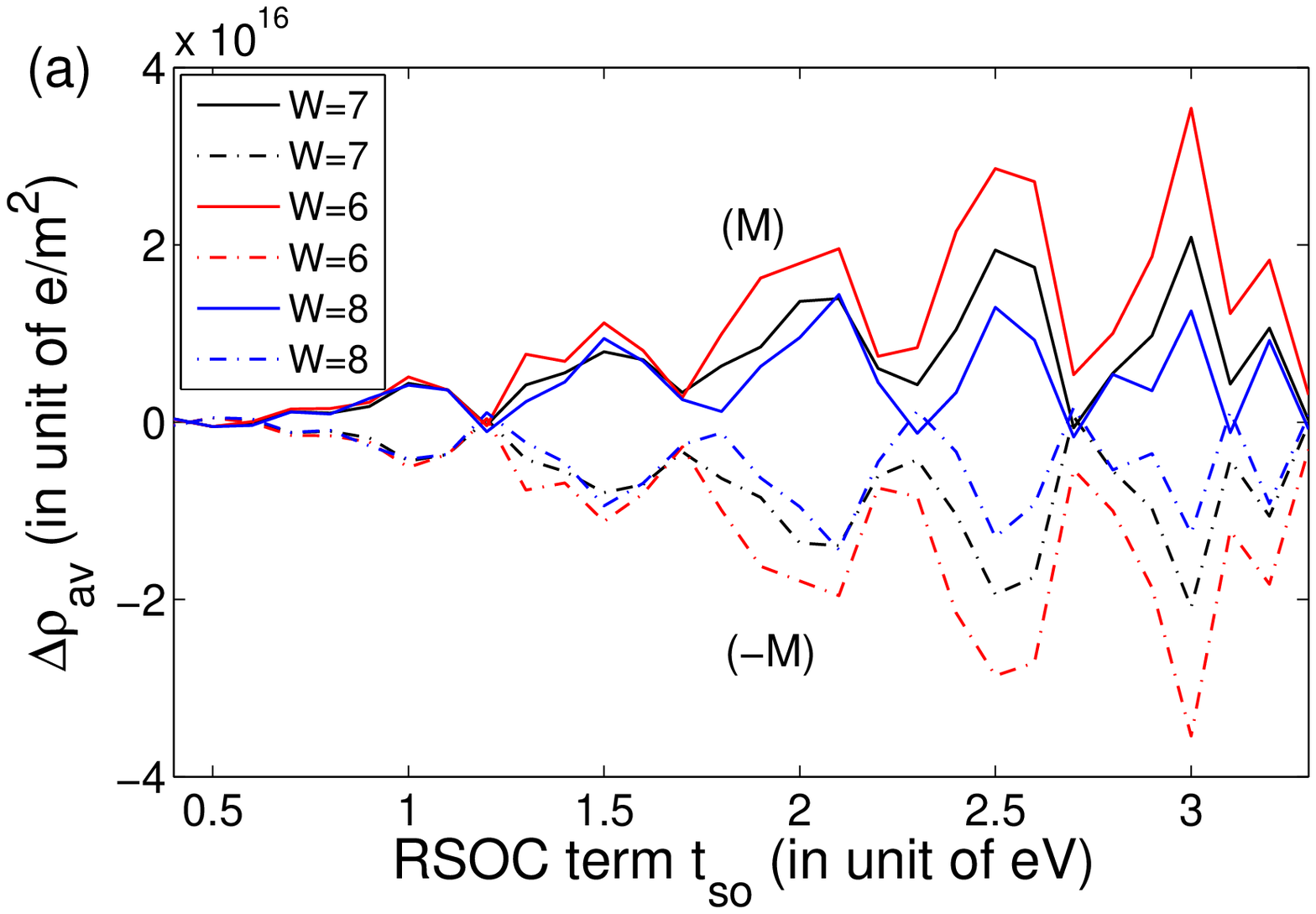}
\includegraphics[width=4.0cm]{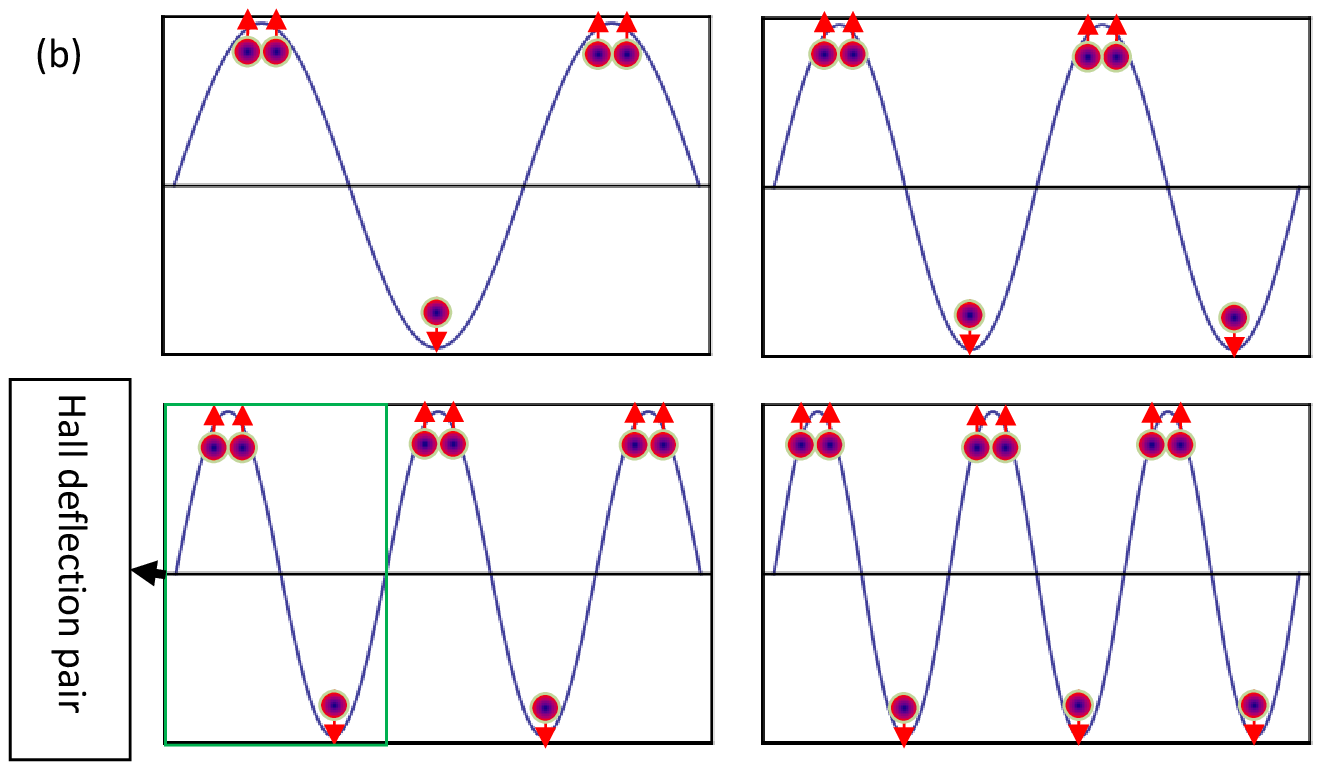}
\caption{(a) The oscillatory increase of $\Delta\rho_{av}$ with the RSOC strength $t_{so}$, for different edge width $W$ and FM moment orientations (solid lines for $z$, dashed lines for $-z$ ). The central region is discretized into a $(200\times100)$ lattice, with dimensions (9 nm$\times$4.5 nm). The $s$-$d$ coupling strength is $M=0.85$ eV, while the bias voltage is $V=4$ V. (b) The schematic diagram of the spin electron distribution due to Yang-Mills-like Lorentz force. The number of Hall deflection pairs increases with increasing RSOC strength.}
\label{fige5}
\end{figure}

Fig. \ref{fige5}(a) shows the dependence of $\Delta\rho_{av}$ on the RSOC strength $t_{so}$ for different edge width $W$. $\Delta\rho_{av}$, and hence the Hall voltage $V_{t}$ across the central region, show an increasing trend with the RSOC strength $t_{so}$, but in an oscillatory manner. The physics underlying this oscillatory increase can be understood in terms of the Yang-Mills-like Lorentz force arising from the RSOC gauge \cite{Tan,Shen}. Here, the FM moments $M$ merely play the role of sustaining a vertical spin polarization of current but do not contribute directly to the Lorentz force. The Lorentz force leads to the transverse separation of electrons of opposite spins [shown schematically in Fig. \ref{fige5}(b)], which we term as ``Hall deflection pair". Since there are unequal number of charges on the two transverse sides, each Hall deflection pair will contribute to a charge Hall voltage. For a fixed $M$, an increase in the RSOC strength results in an increasing number of Hall deflection pairs along the length of the device, as shown in Fig. \ref{fige5}(b). The charge imbalance in a Hall deflection pair coupled with the increase in the number of such pairs with the RSOC strength provide a heuristic explanation of the oscillatory increase of $\Delta\rho_{av}$ and the Hall voltage with the RSOC strength. The sign of $\Delta\rho_{av}$ can be reversed by switching the orientation of FM moment between $\pm z$. Furthermore, $\Delta\rho_{av}$ is not sensitive to the definition of edge, i.e. the general trend remains unchanged for the range of edge width $W$ considered in our calculation.

Previous work on RSOC in semiconductors has shown that when the surface charge density difference is in the order of $10^{12}e/\mathrm{m}^2$, the generated Hall voltage will be large enough for detection (0.1 mV) \cite{Li}. The electron density in our metallic FM RSOC device will be much higher than that in semiconductors, so that $\Delta\rho_{av}$ could attain a value of the order of $10^{16}e/\mathrm{m}^2$ and generate a sizable Hall voltage of $V_t\approx1V$. By selecting optimal $t_{so}$ which corresponds to the peak Hall voltage values (see Fig. \ref{fige5}), we conjecture that a reasonably large $\Delta\rho_{av}$, hence $V_t$ can be measured when the FM moments are oriented along the out-of-plane $z$ axis. The charge density difference $\Delta\rho_{av}$ switches in sign upon reversal of the FM moments to the $-z$ direction. The resulting large difference in the Hall voltage corresponding to the two FM orientations ($\pm z$) suggests that the ISHE can be utilized in a metal FM RSOC system for the sensitive detection of the FM moment orientation.

In summary,we have investigated the inverse spin Hall effect (ISHE) which is induced by the combination of RSOC and $s$-$d$ coupling to the FM moments. A Hall voltage is generated when the FM moments are oriented in the perpendicular-to-plane direction. The Hall voltage increases in an oscillating manner with the RSOC strength $t_{so}$. The polarity of the Hall voltage is reversed when the FM moment is switched to the opposite direction. This property suggests the utility of the ISHE in FM metals with strong RSOC effect for the detection of the FM moment direction, e.g., as a possible memory readback mechanism.

\begin{acknowledgments}
This work was supported by the ASTAR SERC Grant No.
092 101 0060 (R-398-000-061-331) and the NSFC Grant No. 50831002,
51071022.
\end{acknowledgments}


\begin{thebibliography}{10}%
\makeatletter
\providecommand \@ifxundefined [1]{%
 \ifx #1\undefined \expandafter \@firstoftwo
 \else \expandafter \@secondoftwo
\fi
}%
\providecommand \@ifnum [1]{%
 \ifnum #1\expandafter \@firstoftwo
 \else \expandafter \@secondoftwo
\fi
}%
\providecommand \enquote [1]{``#1''}%
\providecommand \bibnamefont  [1]{#1}%
\providecommand \bibfnamefont [1]{#1}%
\providecommand \citenamefont [1]{#1}%
\providecommand\href[0]{\@sanitize\@href}%
\providecommand\@href[1]{\endgroup\@@startlink{#1}\endgroup\@@href}%
\providecommand\@@href[1]{#1\@@endlink}%
\providecommand \@sanitize [0]{\begingroup\catcode`\&12\catcode`\#12\relax}%
\@ifxundefined \pdfoutput {\@firstoftwo}{%
 \@ifnum{\z@=\pdfoutput}{\@firstoftwo}{\@secondoftwo}%
}{%
 \providecommand\@@startlink[1]{\leavevmode}%
 \providecommand\@@endlink[0]{}%
}{%
 \providecommand\@@startlink[1]{%
  \leavevmode
  \pdfstartlink
   attr{/Border[0 0 1 ]/H/I/C[0 1 1]}%
   user{/Subtype/Link/A<</Type/Action/S/URI/URI(#1)>>}%
  \relax
 }%
 \providecommand\@@endlink[0]{\pdfendlink}%
}%
\providecommand \url  [0]{\begingroup\@sanitize \@url }%
\providecommand \@url [1]{\endgroup\@href {#1}{\urlprefix}}%
\providecommand \urlprefix [0]{URL }%
\providecommand \Eprint[0]{\href }%
\@ifxundefined \urlstyle {%
  \providecommand \doi [1]{doi:\discretionary{}{}{}#1}%
}{%
  \providecommand \doi [0]{doi:\discretionary{}{}{}\begingroup
  \urlstyle{rm}\Url }%
}%
\providecommand \doibase [0]{http://dx.doi.org/}%
\providecommand \Doi[1]{\href{\doibase#1}}%
\providecommand \selectlanguage [0]{\@gobble}%
\providecommand \bibinfo [0]{\@secondoftwo}%
\providecommand \bibfield [0]{\@secondoftwo}%
\providecommand \translation [1]{[#1]}%
\providecommand \BibitemOpen[0]{}%
\providecommand \bibitemStop [0]{}%
\providecommand \bibitemNoStop [0]{.\EOS\space}%
\providecommand \EOS [0]{\spacefactor3000\relax}%
\providecommand \BibitemShut [1]{\csname bibitem#1\endcsname}%
\bibitem{Gambardella}%
  \BibitemOpen
  \bibfield{author}{%
  \bibinfo {author} {\bibfnamefont{P.}~\bibnamefont{Gambardella}}, \bibinfo
  {author} {\bibfnamefont{S.}~\bibnamefont{Rusponi}}, \bibinfo {author}
  {\bibfnamefont{M.}~\bibnamefont{Veronese}}, \bibinfo {author}
  {\bibfnamefont{S.~S.}\ \bibnamefont{Dhesi}}, \bibinfo {author}
  {\bibfnamefont{C.}~\bibnamefont{Grazioli}}, \bibinfo {author}
  {\bibfnamefont{A.}~\bibnamefont{Dallmeyer}}, \bibinfo {author}
  {\bibfnamefont{I.}~\bibnamefont{Cabria}}, \bibinfo {author}
  {\bibfnamefont{R.}~\bibnamefont{Zeller}}, \bibinfo {author}
  {\bibfnamefont{P.~H.}\ \bibnamefont{Dederichs}}, \bibinfo {author}
  {\bibfnamefont{K.}~\bibnamefont{Kern}}, \bibinfo {author}
  {\bibfnamefont{C.}~\bibnamefont{Carbone}},\ and\ \bibinfo {author}
  {\bibfnamefont{H.}~\bibnamefont{Brune}},\ }%
  \bibfield{journal}{%
  \bibinfo {journal} {Science}\ }%
  \textbf{\bibinfo {volume} {300}},\ \bibinfo {pages} {1130} (\bibinfo {year}
  {2003})\BibitemShut{NoStop}%
\bibitem{Ast}%
  \BibitemOpen
  \bibfield{author}{%
  \bibinfo {author} {\bibfnamefont{C.~R.}\ \bibnamefont{Ast}}, \bibinfo
  {author} {\bibfnamefont{J.}~\bibnamefont{Henk}}, \bibinfo {author}
  {\bibfnamefont{A.}~\bibnamefont{Ernst}}, \bibinfo {author}
  {\bibfnamefont{L.}~\bibnamefont{Moreschini}}, \bibinfo {author}
  {\bibfnamefont{M.~C.}\ \bibnamefont{Falub}}, \bibinfo {author}
  {\bibfnamefont{D.}~\bibnamefont{Pacil\'e}}, \bibinfo {author}
  {\bibfnamefont{P.}~\bibnamefont{Bruno}}, \bibinfo {author}
  {\bibfnamefont{K.}~\bibnamefont{Kern}},\ and\ \bibinfo {author}
  {\bibfnamefont{M.}~\bibnamefont{Grioni}},\ }%
  \bibfield{journal}{%
  \bibinfo {journal} {Phys. Rev. Lett.}\ }%
  \textbf{\bibinfo {volume} {98}},\ \bibinfo {pages} {186807} (\bibinfo {year}
  {2007})\BibitemShut{NoStop}%
\bibitem{LaShell}%
  \BibitemOpen
  \bibfield{author}{%
  \bibinfo {author} {\bibfnamefont{S.}~\bibnamefont{LaShell}}, \bibinfo
  {author} {\bibfnamefont{B.~A.}\ \bibnamefont{McDougall}},\ and\ \bibinfo
  {author} {\bibfnamefont{E.}~\bibnamefont{Jensen}},\ }%
  \bibfield{journal}{%
  \bibinfo {journal} {Phys. Rev. Lett.}\ }%
  \textbf{\bibinfo {volume} {77}},\ \bibinfo {pages} {3419} (\bibinfo {year}
  {1996})\BibitemShut{NoStop}%
\bibitem{Miron}%
  \BibitemOpen
  \bibfield{author}{%
  \bibinfo {author} {\bibfnamefont{I.~M.}\ \bibnamefont{Miron}}, \bibinfo
  {author} {\bibfnamefont{G.}~\bibnamefont{Gaudin}}, \bibinfo {author}
  {\bibfnamefont{S.}~\bibnamefont{Auffret}}, \bibinfo {author}
  {\bibfnamefont{B.}~\bibnamefont{Rodmacq}}, \bibinfo {author}
  {\bibfnamefont{A.}~\bibnamefont{Schuhl}}, \bibinfo {author}
  {\bibfnamefont{S.}~\bibnamefont{Pizzini}}, \bibinfo {author}
  {\bibfnamefont{J.}~\bibnamefont{Vogel}},\ and\ \bibinfo {author}
  {\bibfnamefont{P.}~\bibnamefont{Gambardella}},\ }%
  \bibfield{journal}{%
  \bibinfo {journal} {Nature Materials}\ }%
  \textbf{\bibinfo {volume} {9}},\ \bibinfo {pages} {230} (\bibinfo {year}
  {2010})\BibitemShut{NoStop}%
\bibitem{saitoh}%
  \BibitemOpen
  \bibfield{author}{%
  \bibinfo {author} {\bibfnamefont{E.}~\bibnamefont{Saitoh}}, \bibinfo {author}
  {\bibfnamefont{M.}~\bibnamefont{Ueda}}, \bibinfo {author}
  {\bibfnamefont{H.}~\bibnamefont{Miyajima}},\ and\ \bibinfo {author}
  {\bibfnamefont{G.}~\bibnamefont{Tatara}},\ }%
  \bibfield{journal}{%
  \bibinfo {journal} {Applied Physics Letters}\ }%
  \textbf{\bibinfo {volume} {88}},\ \bibinfo {pages} {182509} (\bibinfo {year}
  {2006})\BibitemShut{NoStop}%
\bibitem{Hankiewicz}%
  \BibitemOpen
  \bibfield{author}{%
  \bibinfo {author} {\bibfnamefont{E.~M.}\ \bibnamefont{Hankiewicz}}, \bibinfo
  {author} {\bibfnamefont{J.}~\bibnamefont{Li}}, \bibinfo {author}
  {\bibfnamefont{T.}~\bibnamefont{Jungwirth}}, \bibinfo {author}
  {\bibfnamefont{Q.}~\bibnamefont{Niu}}, \bibinfo {author}
  {\bibfnamefont{S.-Q.}\ \bibnamefont{Shen}},\ and\ \bibinfo {author}
  {\bibfnamefont{J.}~\bibnamefont{Sinova}},\ }%
  \bibfield{journal}{%
  \bibinfo {journal} {Phys. Rev. B}\ }%
  \textbf{\bibinfo {volume} {72}},\ \bibinfo {pages} {155305} (\bibinfo {year}
  {2005})\BibitemShut{NoStop}%
\bibitem{Valenzuela}%
  \BibitemOpen
  \bibfield{author}{%
  \bibinfo {author} {\bibfnamefont{S.~O.}\ \bibnamefont{Valenzuela}}\ and\
  \bibinfo {author} {\bibfnamefont{M.}~\bibnamefont{Tinkham}},\ }%
  \bibfield{journal}{%
  \bibinfo {journal} {Nature}\ }%
  \textbf{\bibinfo {volume} {442}},\ \bibinfo {pages} {176} (\bibinfo {year}
  {2006})\BibitemShut{NoStop}%
\bibitem{Zhang}%
  \BibitemOpen
  \bibfield{author}{%
  \bibinfo {author} {\bibfnamefont{J.~J.}\ \bibnamefont{Zhang}}, \bibinfo
  {author} {\bibfnamefont{F.}~\bibnamefont{Liang}},\ and\ \bibinfo {author}
  {\bibfnamefont{J.}~\bibnamefont{Wang}},\ }%
  \bibfield{journal}{%
  \bibinfo {journal} {Eur. Phys. J. B}\ }%
  \textbf{\bibinfo {volume} {72}},\ \bibinfo {pages} {105} (\bibinfo {year}
  {2009})\BibitemShut{NoStop}%
\bibitem{Saitoh1}%
  \BibitemOpen
  \bibfield{author}{%
  \bibinfo {author} {\bibfnamefont{E.}~\bibnamefont{Saitoh}}, \bibinfo {author}
  {\bibfnamefont{M.}~\bibnamefont{Ueda}}, \bibinfo {author}
  {\bibfnamefont{H.}~\bibnamefont{Miyajima}},\ and\ \bibinfo {author}
  {\bibfnamefont{G.}~\bibnamefont{Tatara}},\ }%
  \bibfield{journal}{%
  \bibinfo {journal} {Appl. Phys. Lett.}\ }%
  \textbf{\bibinfo {volume} {88}},\ \bibinfo {pages} {182509} (\bibinfo {year}
  {2006})\BibitemShut{NoStop}%
\bibitem{Kimura}%
  \BibitemOpen
  \bibfield{author}{%
  \bibinfo {author} {\bibfnamefont{T.}~\bibnamefont{Kimura}}, \bibinfo {author}
  {\bibfnamefont{Y.}~\bibnamefont{Otani}}, \bibinfo {author}
  {\bibfnamefont{T.}~\bibnamefont{Sato}}, \bibinfo {author}
  {\bibfnamefont{S.}~\bibnamefont{Takahashi}},\ and\ \bibinfo {author}
  {\bibfnamefont{S.}~\bibnamefont{Maekawa}},\ }%
  \bibfield{journal}{%
  \bibinfo {journal} {Phys. Rev. Lett.}\ }%
  \textbf{\bibinfo {volume} {98}},\ \bibinfo {pages} {156601} (\bibinfo {year}
  {2007})\BibitemShut{NoStop}%
\bibitem{Xing}%
  \BibitemOpen
  \bibfield{author}{%
  \bibinfo {author} {\bibfnamefont{Y.}~\bibnamefont{Xing}}, \bibinfo {author}
  {\bibfnamefont{Q.-F.}\ \bibnamefont{Sun}},\ and\ \bibinfo {author}
  {\bibfnamefont{J.}~\bibnamefont{Wang}},\ }%
  \bibfield{journal}{%
  \bibinfo {journal} {Phys. Rev. B}\ }%
  \textbf{\bibinfo {volume} {75}},\ \bibinfo {pages} {075324} (\bibinfo {year}
  {2007})\BibitemShut{NoStop}%
\bibitem{Li}%
  \BibitemOpen
  \bibfield{author}{%
  \bibinfo {author} {\bibfnamefont{J.}~\bibnamefont{Li}}\ and\ \bibinfo
  {author} {\bibfnamefont{S.~Q.}\ \bibnamefont{Shen}},\ }%
  \bibfield{journal}{%
  \bibinfo {journal} {Phys. Rev. B}\ }%
  \textbf{\bibinfo {volume} {76}},\ \bibinfo {pages} {153302} (\bibinfo {year}
  {2007})\BibitemShut{NoStop}%
\bibitem{Ando}%
  \BibitemOpen
  \bibfield{author}{%
  \bibinfo {author} {\bibfnamefont{K.}~\bibnamefont{Ando}}\ and\ \bibinfo
  {author} {\bibfnamefont{E.}~\bibnamefont{Saitoh}},\ }%
  \bibfield{journal}{%
  \bibinfo {journal} {J. Appl. Phys.}\ }%
  \textbf{\bibinfo {volume} {108}},\ \bibinfo {pages} {113925} (\bibinfo {year}
  {2010})\BibitemShut{NoStop}%
\bibitem{Rashba}%
  \BibitemOpen
  \bibfield{author}{%
  \bibinfo {author} {\bibfnamefont{E.~I.}\ \bibnamefont{Rashba}},\ }%
  \bibfield{journal}{%
  \bibinfo {journal} {Phys. Rev. B}\ }%
  \textbf{\bibinfo {volume} {70}},\ \bibinfo {pages} {161201(R)} (\bibinfo
  {year} {2004})\BibitemShut{NoStop}%
\bibitem{Wawrzyniaka}%
  \BibitemOpen
  \bibfield{author}{%
  \bibinfo {author} {\bibfnamefont{M.}~\bibnamefont{Wawrzyniaka}}, \bibinfo
  {author} {\bibfnamefont{M.}~\bibnamefont{Mackowski}}, \bibinfo {author}
  {\bibfnamefont{Z.}~\bibnamefont{Sniadecki}}, \bibinfo {author}
  {\bibfnamefont{B.}~\bibnamefont{Idzikowski}},\ and\ \bibinfo {author}
  {\bibfnamefont{J.}~\bibnamefont{Martinek}},\ }%
  \bibfield{journal}{%
  \bibinfo {journal} {Acta Physica Polonica A.}\ }%
  \textbf{\bibinfo {volume} {118}},\ \bibinfo {pages} {375} (\bibinfo {year}
  {2010})\BibitemShut{NoStop}%
\bibitem{Henk}%
  \BibitemOpen
  \bibfield{author}{%
  \bibinfo {author} {\bibfnamefont{J.}~\bibnamefont{Henk}}, \bibinfo {author}
  {\bibfnamefont{M.}~\bibnamefont{Hoesch}}, \bibinfo {author}
  {\bibfnamefont{J.}~\bibnamefont{Osterwalder}}, \bibinfo {author}
  {\bibfnamefont{A.}~\bibnamefont{Ernst1}},\ and\ \bibinfo {author}
  {\bibfnamefont{P.}~\bibnamefont{Bruno}},\ }%
  \bibfield{journal}{%
  \bibinfo {journal} {J. Phys.: Condens. Matter}\ }%
  \textbf{\bibinfo {volume} {16}},\ \bibinfo {pages} {7581} (\bibinfo {year}
  {2004})\BibitemShut{NoStop}%
\bibitem{Krupin}%
  \BibitemOpen
  \bibfield{author}{%
  \bibinfo {author} {\bibfnamefont{O.}~\bibnamefont{Krupin}}, \bibinfo {author}
  {\bibfnamefont{G.}~\bibnamefont{Bihlmayer}}, \bibinfo {author}
  {\bibfnamefont{K.}~\bibnamefont{Starke}}, \bibinfo {author}
  {\bibfnamefont{S.}~\bibnamefont{Gorovikov}}, \bibinfo {author}
  {\bibfnamefont{J.~E.}\ \bibnamefont{Prieto}}, \bibinfo {author}
  {\bibfnamefont{K.}~\bibnamefont{D\"obrich}}, \bibinfo {author}
  {\bibfnamefont{S.}~\bibnamefont{Bl\"ugel}},\ and\ \bibinfo {author}
  {\bibfnamefont{G.}~\bibnamefont{Kaindl}},\ }%
  \bibfield{journal}{%
  \bibinfo {journal} {Phys. Rev. B}\ }%
  \textbf{\bibinfo {volume} {71}} (\bibinfo {year} {2005})\BibitemShut{NoStop}%
\bibitem{Wakoh}%
  \BibitemOpen
  \bibfield{author}{%
  \bibinfo {author} {\bibfnamefont{S.}~\bibnamefont{Wakoh}}\ and\ \bibinfo
  {author} {\bibfnamefont{J.}~\bibnamefont{Yamashita}},\ }%
  \bibfield{journal}{%
  \bibinfo {journal} {Journal of the Physical Society of Japan}\ }%
  \textbf{\bibinfo {volume} {28}},\ \bibinfo {pages} {1151} (\bibinfo {year}
  {1970})\BibitemShut{NoStop}%
\bibitem{Tan}%
  \BibitemOpen
  \bibfield{author}{%
  \bibinfo {author} {\bibfnamefont{S.~G.}\ \bibnamefont{Tan}}, \bibinfo
  {author} {\bibfnamefont{M.~B.~A.}\ \bibnamefont{Jalil}}, \bibinfo {author}
  {\bibfnamefont{X.-J.}\ \bibnamefont{Liu}},\ and\ \bibinfo {author}
  {\bibfnamefont{T.}~\bibnamefont{Fujita}},\ }%
  \bibfield{journal}{%
  \bibinfo {journal} {Phys. Rev. B.}\ }%
  \textbf{\bibinfo {volume} {78}},\ \bibinfo {pages} {245321} (\bibinfo {year}
  {2008})\BibitemShut{NoStop}%
\bibitem{Shen}%
  \BibitemOpen
  \bibfield{author}{%
  \bibinfo {author} {\bibfnamefont{S.-Q.}\ \bibnamefont{Shen}},\ }%
  \bibfield{journal}{%
  \bibinfo {journal} {Phys. Rev. Lett.}\ }%
  \textbf{\bibinfo {volume} {95}},\ \bibinfo {pages} {187203} (\bibinfo {year}
  {2005})\BibitemShut{NoStop}%
\end{thebibliography}
%
\end{document}